# Fresnel Coefficients, coherent optical scattering, and planar waveguiding

*Alejandro Doval, Carlos Damián Rodríguez-Fernández, Héctor González-Núñez and Raúl de la Fuente\**

*Grupo de Nanomateriais, Fotónica e Materia Branda, Departamento de Física Aplicada, Universidade de Santiago de Compostela, E-15782, Santiago de Compostela, Spain.*

*\*Corresponding autor: raul.delafuente@usc.es*

**Abstract.** One-dimensional optical waveguiding is revisited using the electromagnetic deduction of Fresnel formulas relating the incident, reflected, and transmitted waves on the abrupt interface between two different optical media. Throughout the paper, different optical configurations are analyzed from this perspective revealing that, in all cases, those solutions corresponding to waveguiding configurations share common features. These shared characteristics are: firstly, the guided modes always emerge as singular solutions of the associate coherent optical scattering problem and, secondly, the resonance condition that must be satisfied corresponds to the pole of the associated Fresnel Coefficients.

## 1. Introduction

Throughout the centuries, numerous scientists made notable contributions to the field of optics. Among them, an undoubtedly outstanding figure is that of Augustin-Jean Fresnel. Despite his short and sickly life, Fresnel is well known not only for his plentiful and varied contributions, but also for giving a definitive push to the wave theory of light, in opposition to Newton's corpuscular theory. In this regard, it is worth of special mention his work on the theory of diffraction, on the transverse nature and polarization of light waves, and on the birefringence of crystals [1]. Striking evidence of his scientific heritage is the number of concepts and elements bearing his name. For example, there are Fresnel integrals and diffraction regimes, but also Fresnel lenses, lanterns, or zone plates.

Among all the topics he dealt with along his life, a particularly relevant one is that of the incidence of light on a flat surface separating two transparent media of different refractive indices. In his treatment, he derived the law of sines and tangents relating the perpendicular and tangential components of the reflected waves in relation to those of the incident waves [2]. These relationships, together with those for the transmitted waves, are currently known as the Fresnel Equations or Fresnel Coefficients. It is noteworthy to mention that this work preceded Maxwell's electromagnetic theory, being, one of the most brilliant contributions related to the transverse nature of electromagnetic waves. In fact, from the perspective of this subsequent theory, these pioneering equations arise as a natural consequence of the boundary conditions of the electromagnetic fields on a surface producing an electric and/or magnetic discontinuity.

Fresnel's approach was historically used not only to analyze the boundary problem between two media of different refractive indices, but also to study a large number of other problems in optics, such as those related to multiple parallel interfaces [3]. Interestingly, there are certain optical configurations and optical problems in which particular boundary conditions need to be defined or specified and whose traditional solution does not involve the Fresnel relations at all. As far as we can see, the usage of problem-dependent approaches in those situations where the Fresnel formalism can be naturally applied is misleading. Solutions applying Fresnel relations allow to visualize the common physical basis to these boundary phenomena.

In this work, we apply the Fresnel formalism to the study of electromagnetic waveguiding, revealing that guided modes correspond to singular solutions of a boundary problem in which Fresnel Coefficients tend to infinity. In the following sections, we consider different waveguiding configurations involving two or three media with different permittivities varying in a given direction, or a medium in which the permittivity varies continuously. We will refer as coherent optical scattering to these cases of study, in which a monochromatic optical signal arrives to a general optical system and gives rise to two waves, one reflected and one transmitted. Note that, in contrast to other scattering problems in physics, the vector nature of EM waves imposes the need to distinguish between two types of waves in this sort of problems: electric and magnetic transverse waves. The study of the presented situations using the Fresnel Coefficients allows us to obtain a more general vision of the physics associated to these boundary problems and to extract general conclusions that often pass unnoticed.

## 2. Coherent Optical Scattering

Fresnel's approach to a boundary problem can be applied to solve whatever problem involving coherent optical scattering, i.e., that in which a coherent optical signal propagates in a region defined by a given optical potential. This potential could be either determined by the variation of the refractive index, n, or its square, the dielectric permittivity, $\varepsilon = n^2$. For convenience, we define these potentials as: $-(\omega/c)^2 \varepsilon$, with $\omega$ the frequency of the optical signal, and c the speed of light in vacuum (the minus sign has been chosen in analogy with the quantum mechanics equivalent problem [4]). In the case of transparent media, we simply have a potential which is proportional to the dielectric constant, while in the case of absorbing media, we have, in addition, an average energy dissipation associated with the imaginary part of the permittivity.

In Fig. 1 we consider some examples. In the first two, we show step potentials, which are those associated with a discontinuity in the permittivity. In these cases, the tangential component of the wavevector is a conserved quantity (hereafter referred to the wave propagation constant, $\beta$) and different solutions are distinguished according to its value. For instance, when we consider a transition from a region of lower potential to other of higher potential in transparent media, the phenomenon of total reflection arises if this conserved quantity exceeds the wavenumber of the wave in the second medium, that means, $n_2\omega/c < \beta < n_1\omega/c$. Meanwhile, $\beta < n_2\omega/c$ (or $\beta < n_1\omega/c$ in Fig.1a), corresponds to normal refraction. Furthermore, surface plasmon generation takes place in the scattering configuration of Fig. 1a if $-\varepsilon_2 > \varepsilon_1 > 0$.

In the case of considering three media of different but constant permittivity, we can find both rectangular potential barriers (Fig. 1c) or wells (Fig. 1d) and, hence, total reflection, frustrated total reflection or waveguiding. As before, radiation modes correspond to waves that propagate with propagation constants lower than any of the wavenumbers. In contrast, total reflection occurs in Fig. 1c for higher values of $\beta$ and it can be frustrated or not depending on the thickness of the second medium. Besides, in Fig. 1d waveguiding takes places if $n_1\omega/c$, $n_2\omega/c < \beta < n_2\omega/c$. Finally, in Fig. 1e, we show a gradual variation of the refractive index within a medium. Since the potential is asymmetric, we can distinguish between radiation, semi-radiation (totally reflected) and guided modes. The grey zone in Figs. 1d and 1e corresponds to possible discrete values of the squared propagation constant of the guided modes.

The results reported in the previous paragraph are well known in the field and their deduction can be found in most optics textbooks (see for example Refs. [5] and [6]). In this kind of optics manuals, the solutions given to these guiding configurations are commonly obtained from particular approaches that are specific for each configuration, hiding part of the common phenomenology of these boundary problems. For this reason, on the following sections, we are going to use a unified approach, the conceptual framework provided by the Fresnel Equations, to perform a common analysis of some of the waveguiding configurations presented up to the moment. We will see that, under this strategy, the guided modes always arise as singular solutions of the scattering problem or, equivalently, as poles of the total Fresnel Coefficients (that means, the Fresnel Coefficients of the complete system). However, prior to that, we would like to perform a brief review of the formalism that gives rise to the Fresnel Equations.

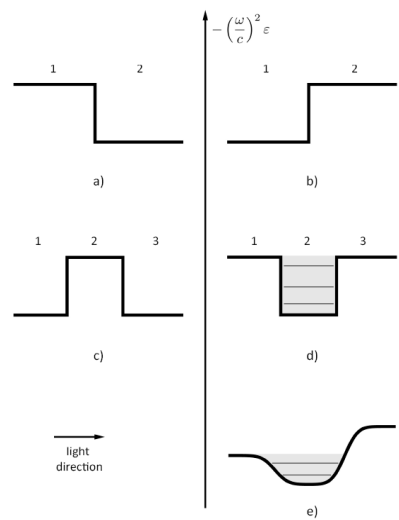

Figure 1. Different configurations of coherent optical scattering problems. Negative (a) and positive (b) step potential; rectangular barrier (c) and well (d); general well (e). The numbers refer to media of different permittivities and continuous lines represents discrete guided modes

## 3. The Fresnel Equations

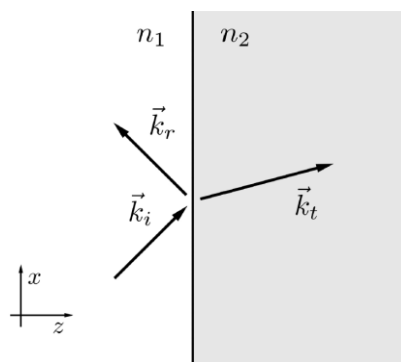

Figure 2. General schema of coherent optical scattering in an abrupt planar interface. The plane of the figure corresponds to the plane of incidence, defined by the incident wave propagation direction and the vector normal to the boundary surface, $\vec{z}$

The Fresnel Equations describe the relationship between complex wave amplitudes of reflected, transmitted and incident electromagnetic waves at the boundary of two different optical media within the framework of classical electromagnetic theory. As shown in Fig. 2, let us consider an electromagnetic wave of frequency ω travelling in a medium of complex refractive index $n_1$ and arriving at a certain angle at the flat boundary surface with a medium of complex refractive index $n_2$. This incoming electromagnetic wave gives rise to a wave reflected back to medium 1 and another one transmitted or refracted towards medium 2.

The relationship between these waves is defined by the boundary conditions of the EM waves, i.e., the tangential or parallel components of both electric and magnetic fields on the boundary surface must be continuous [7], condition that keeps the frequency of the three waves unchanged at both sides of the frontier. In order to apply boundary conditions to the problem, it is convenient to distinguish between two types of waves: Transverse Electric waves (also known as TE or s-polarized), in which the electric field vibrates perpendicular to the plane of incidence, and Transverse Magnetic waves (also known as TM or p-polarized), in which the vibration perpendicular to the plane of incidence is that of the magnetic field. For simplicity, we are going to assume that the electromagnetic waves upon consideration are plane waves, which are the simplest solution to Maxwell's equations. Upon this assumption, the electric field $\vec{E}_j$ of the incident ($j = i$), reflected ($j = r$) and transmitted ($j = t$) waves involved in our boundary problem are:

$$\vec{E}_j = A_j \vec{e}_j e^{i\varphi_j} \quad j = i, r, t$$
$$\varphi_j = \vec{k}_j \vec{r} - \omega t \tag{1}$$

where $A_j$ is the amplitude of the electric field $\vec{E}_j$ which vibrates in the direction of the unit vector $\vec{e}_j$ (perpendicular to the plane of incidence for TE waves and parallel to the plane of incidence for TM waves). On the other hand, $\varphi_j$ is its phase, which depends on both the light frequency ω and the wavevector of light in the material, $\vec{k}_j$. These wavevectors are, in general, complex numbers with their real part associated to the phase velocity, and their imaginary part associated with absorption in the medium. The modulus of the wavevector (the wavenumber) is related to the frequency of the waves and the refractive index of the medium as follows:

$$k_j = \left|\vec{k}_j\right| = \frac{\omega}{c}\left|n_j\right| \quad n_i, n_r = n_1; \ n_t = n_2 \tag{2}$$

Considering that the plane of incidence is perpendicular to the y-axis, and that the normal vector of the boundary surface is directed along the z-axis, the continuity of the field component tangential to the plane of incidence of the waves given by the boundary condition is easily written as:

$$k_{ix} = k_{rx} = k_{tx} = \beta \tag{3}$$

which defines the wave propagation constant, $\beta$. Meanwhile, the components normal to the boundary surface are defined by Eqs. (2) and (3):

$$k_{iz} = -k_{rz}^* = \sqrt{(\omega n_1/c)^2 - \beta^2}$$
$$k_{tz} = \sqrt{(\omega n_2/c)^2 - \beta^2} \tag{4}$$

where the change of sign in the real part of the wavevector takes into account that the incident wave is an "incoming" wave in the second medium, while the reflected wave is an "outgoing" wave to the first medium. Furthermore, since the value of the fields must be finite at $z \to -\infty$, the imaginary part of the normal components of the wave vector must have the same sign for both incident and reflected waves. On the other hand, the boundary conditions imply two relationships between the electric field amplitudes of the three waves as [7]:

$$(A_i + A_r) = \eta A_t$$
$$(A_i - A_r)k_{iz} = A_t k_{tz}/\eta \quad (5)$$

where η = 1 for TE waves and η ≡ n = $n_2/n_1$ for TM waves, being n the relative refractive index of medium 2 with respect to medium 1. These two expressions mean that for TE (TM) waves, both the tangential electric (magnetic) field and its derivative are continuous at the boundary. These conditions can be recast in matrix form as follows:

$$M_{TJ}\begin{pmatrix} A_r \\ A_t \end{pmatrix} = \begin{pmatrix} -1 & \eta \\ k_{iz} & k_{tz}/\eta \end{pmatrix}\begin{pmatrix} A_r \\ A_t \end{pmatrix} = A_i\begin{pmatrix} 1 \\ k_{iz} \end{pmatrix} \quad J = E, M \quad (6)$$

Multiplying by the inverse matrix, $M_{TJ}^{-1}$, we obtain the ratio between the amplitudes of the reflected and transmitted waves and the amplitude of the incident wave:

$$\begin{pmatrix} A_r \\ A_t \end{pmatrix} = \begin{pmatrix} r_{TJ} \\ t_{TJ} \end{pmatrix} A_i = A_i M_{TJ}^{-1}\begin{pmatrix} 1 \\ k_{iz} \end{pmatrix} \quad J = E, M \quad (7)$$

being the $r_{TJ}$ and $t_{TJ}$ parameters nothing more than the well-known Fresnel Coefficients:

$$r_{TE} = \frac{k_{iz} - k_{tz}}{k_{iz} + k_{tz}}; \quad t_{TE} = \frac{2k_{iz}}{k_{iz} + k_{tz}}$$
$$r_{TM} = \frac{nk_{iz} - k_{tz}/n}{nk_{iz} + k_{tz}/n}; \quad t_{TM} = \frac{2k_{iz}}{nk_{iz} + k_{tz}/n} \quad (8)$$

either, expressed as a function of the dielectric constants of both media, $\varepsilon_j = n_j^2, j = 1, 2$, yield:

$$r_{TE} = \frac{k_{iz} - k_{tz}}{k_{iz} + k_{tz}}; \quad t_{TE} = \frac{2k_{iz}}{k_{iz} + k_{tz}}$$
$$r_{TM} = \frac{\varepsilon_2 k_{iz} - \varepsilon_1 k_{tz}}{\varepsilon_2 k_{iz} + \varepsilon_1 k_{tz}}; \quad t_{TM} = \frac{2\sqrt{\varepsilon_1\varepsilon_2}k_{iz}}{\varepsilon_2 k_{iz} + \varepsilon_1 k_{tz}} \quad (9)$$

Note that, in transparent media, these formulas are usually written in terms of the angles of incidence and refraction, but this alternative form is not useful here.

**4. Surface plasmons**

The Fresnel Equations are often used to obtain insightful descriptions of classical optical problems involving dielectric media such as the existence of Brewster's angle or the phenomenon of total reflection, if $n_1 > n_2$. However, their validity is not limited to dielectric media, but it also holds in conductor materials. In this section, we will consider a problem that is not normally addressed using the Fresnel Equations, the generation of surface plasmons [8].

In order to face the plasmon theory with the tools of the Fresnel Equations, we have to describe plasmons in the very same language of the Fresnel formalism. Hence, the question we want to address is: in view of the results presented in the previous section, can light be guided on a boundary? A waveguide can be described as an optical system that allows the propagation of light in a bounded region without an external source, hence $A_i = 0$ in Eqs. (5-7). At first sight, just by considering Eq. (7), the solution of our problem is the trivial solution $A_r = A_t = 0$, that is, no light. However, there is another solution for the problem, which corresponds to the situation where the next condition is satisfied:

$$\det[M_{TJ}] = -(k_{tz}/\eta + k_{iz}\eta) = 0 \tag{10}$$

In this situation, Eq. (7) does not hold anymore since there is no inverse matrix for $M_{TJ}$ because its determinant is null. Hence, we must address the problem directly from Eq. (6). For TE waves, Eq. (6) together with Eq. (10) yields $k_{tz} = -k_{iz}$, which is a non-significant solution. Nevertheless, the situation is by far different when considering TM waves. For those we get:

$$k_{tz}/n - k_{rz}^* n = 0 \tag{11}$$

where the expression was written as a function of the wavevector of the reflected wave because we are considering the case of a null incident wave at the boundary. This equation can be rearranged to be expressed as a function of the dielectric constant:

$$-\varepsilon_2 k_{rz}^* + \varepsilon_1 k_{tz} = 0 \tag{12}$$

which corresponds to the well-known plasmon resonance condition [8] (note that this equation is usually written in terms of the normal component of the incident wavevector: $k_{iz} = -k_{rz}^*$). Combining Eq. (12) and Eq. (4) together and working out the resulting expression, the formula of the mode propagation constant is obtained:

$$\beta = \frac{\omega}{c}\sqrt{\frac{\varepsilon_1 \varepsilon_2}{\varepsilon_1 + \varepsilon_2}} \tag{13}$$

as well as the relationship between the transverse and tangential components of the wave vectors:

$$\sqrt{\frac{\varepsilon_1}{\varepsilon_2}} k_{tz} = -\sqrt{\frac{\varepsilon_2}{\varepsilon_1}} k_{rz}^* = \beta \tag{14}$$

The simplest solution of Eqs. (12) and (13) corresponds to electric permittivity of opposite sign. For example, considering medium 2 with negative permittivity we would have:

$$-\varepsilon_2 > \varepsilon_1 > 0 \tag{15}$$

which means that $\beta$ would be real and, $k_{rz}$ and $k_{tz}$, pure imaginary. In a general situation, this condition corresponds to the incidence of a wave from a transparent dielectric medium to a conductive medium with a frequency lower than the plasma frequency of the conductor, and the imaginary part of the permittivity approaching to zero. In terms of the refractive indices of the media, this situation is written as:

$$n_j = n'_j + in''_j \quad j = 1, 2$$
$$n'_1 = \sqrt{\varepsilon_1}$$
$$n''_2 = -i\sqrt{\varepsilon_2} = \sqrt{-\varepsilon_2} \tag{16}$$

Henceforth, substituting these expressions in Eqs. (14) and (13), $k_{rz}$ and $k_{tz}$ can be rewritten as:

$$k_{tz} = i\frac{n''^2_2}{\gamma}\frac{\omega}{c}, \quad k_{rz} = i\frac{n'^2_1}{\gamma}\frac{\omega}{c}, \quad \gamma = \sqrt{n''^2_2 - n'^2_1} \tag{17}$$

Since the z-component of the wavevectors are purely imaginary, the wave is evanescent at both sides of the interface, as seen in Fig. 3. This electromagnetic wave is strongly coupled to the charge motion at the metal surface (the surface plasmon) and, as a whole, they give rise to a surface plasmon polariton (SPP).

Finally, going back to the matrix from Eq. (6), in the case of $A_i = 0$, the relation between amplitudes for the TM waves is:

$$n_1 A_r = n_2 A_t. \tag{18}$$

These results can be summarized in the following final expressions for electric field amplitudes:

$$E_z = A e^{i\beta z} \begin{cases} n_2'' e^{n_1'^2 \omega z / \gamma c} & z < 0 \\ i n_1' e^{-n_2''^2 \omega z / \gamma c} & z \geq 0 \end{cases}$$

$$E_x = E_z \begin{cases} -i n_1' / n_2'' & z < 0 \\ i n_2'' / n_1' & z \geq 0 \end{cases} \tag{19}$$

where A is an arbitrary constant.

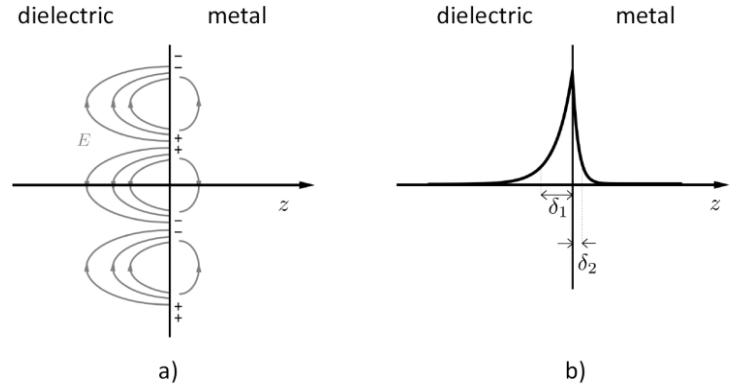

Figure 3. Plot of the electric fields of charges (a) and wave (b) associated to the SPP. $\delta_1, \delta_2$ represent the skin depth in the dielectric and metal media, respectively.

We would like to emphasize that the results reported here to model plasmons are not new, but the novelty arises from the way of retrieving them using the Fresnel formalism. Within this framework, the plasmon resonance condition arises naturally by modelling the typical optical coherent scattering problem of incidence of light at an abrupt interface.

Before ending this section, two final remarks must be highlighted. First, despite we extracted the resonance condition from the Eq. (6), it could be directly obtained from Eq. (5) by setting $A_i = 0$. However, in our opinion, this way of proceeding is better to visualize the singular nature of the resonance condition. Second, from the expression of the Fresnel Coefficients, it should be noted that the denominator of the TM coefficients corresponds to minus the determinant of the $M_{TJ}$ matrix. For this reason, the resonance condition of Eq. (10) is verified if:

$$r_{12} = t_{12} \to \infty, \tag{20}$$

which is a natural observation since it is precisely for the plasmonic modes that a division by zero is made to obtain the Fresnel Coefficients. Hence, it means that this simple form of optical guiding corresponds to the pole of the Fresnel formulas. It must be mentioned that the infiniteness of the reflection Fresnel Coefficient for plasmonic modes was already noted by Cardona [9] for the first time using the Fresnel law of tangents to express the reflection coefficient. Surprisingly, the same condition for the transmission coefficient was not mentioned in that work.

## 5. Step-index planar waveguides

In this section, we study the propagation of light in a step-index planar waveguide using the Fresnel formalism developed before. We begin with a system composed of three media, as shown in Fig. 4, each one with its own refractive index, and planar interfaces as separation boundaries between them. Light impinges in the system from medium 1, afterwards, one part is reflected backwards, another part gets trapped (or not) in medium 2, which has a characteristic thickness *d*, and a third part leaves the system towards medium 3.

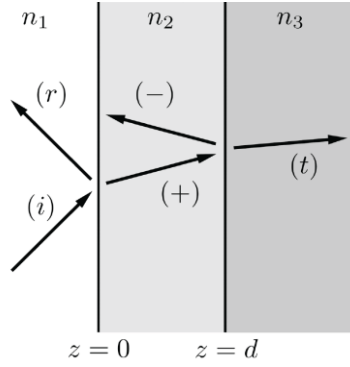

Fig. 4. Coherent optical scattering at two abrupt interfaces.

The strategy we follow for addressing the system is that used for studying, for instance, a Fabry-Perot interferometer [10], and can be generalized to model an arbitrary multilayer with flat boundaries [11]. We denote as $A_+$ the amplitude of the wave travelling through medium 2 towards medium 3 and we denote as $A_-$ the wave propagating through medium 2 towards medium 1. At the plane z = 0, there are two "incoming" waves, (i) and (-), and two "outcoming" waves, (r) and (+). The relationship between the amplitudes of these waves can be obtained from the definition of the Fresnel Coefficients and the application of the superposition principle:

$$A_+ = t_{12}A_i + r_{21}A_- \\ A_r = r_{12}A_i + t_{21}A_- \quad (21)$$

where the subindices "jk" mean that the wave is incident from medium j towards medium k. Meanwhile, the relation between the waves that get trapped in medium 2 can be found by performing one roundtrip through medium 2 beginning and ending at z = 0 (see Fig. 5):

$$A_- e^{i\beta x} = r_{23} A_+ e^{i\beta(x - 2\Delta x)} e^{i2\vec{k}_2 \Delta \vec{r}} \\ \Delta \vec{r} = \Delta x \hat{x} + d\hat{z} \quad (22)$$

or simplifying:

$$A_- = r_{23} A_+ e^{i\phi}; \quad \phi = 2k_{2z}d \\ k_{2z} = \sqrt{(\omega n_2/c)^2 - \beta^2} \quad (23)$$

Being $k_{\pm z} = \pm k_{2z}$ the normal components of the waves travelling within medium 2. The relationship between the transmitted wave and the others can be extracted on a similar way. Observing Fig. (4), the wave transmitted at the observation plane z = d can be written as:

$$A_t = t_{23} A_+ e^{i\phi/2} \quad (24)$$

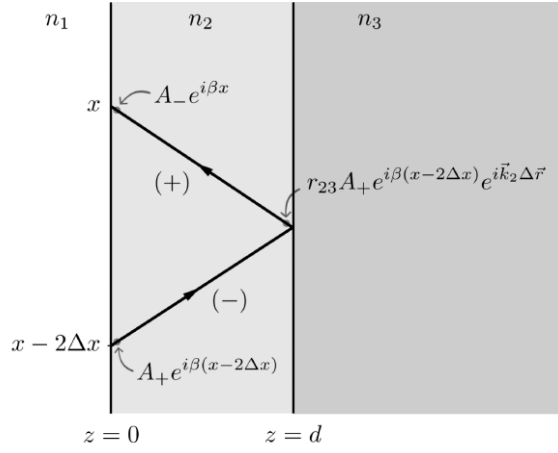

Figure 5. Plot of wave propagation to illustrate deduction of Eq. (22)

The relationships between the different amplitudes are summarized in Eqs. (21), (23) and (24). We first consider the case where there is an external wave, i.e., $A_i \neq 0$. In this situation, the amplitude of all other waves can be written as a function of the amplitude of this incident wave. After some algebra, it is obtained:

$$A_+ = \frac{t_{12}}{1 - r_{21}r_{23}e^{i\phi}} A_i$$

$$A_r = r_{123} A_i = \left[ r_{12} + \frac{t_{21}t_{12}r_{23}e^{i\phi}}{1 - r_{21}r_{23}e^{i\phi}} \right] A_i = \frac{-r_{21} + r_{23}e^{i\phi}}{1 - r_{21}r_{23}e^{i\phi}} A_i$$

$$A_- = \frac{t_{12}r_{23}e^{i\phi}}{1 - r_{21}r_{23}e^{i\phi}} A_i$$

$$A_t = t_{123} A_i = \frac{t_{12}t_{23}e^{i\phi/2}}{1 - r_{21}r_{23}e^{i\phi}} A_i$$

(25)

where the Stokes relations [5] were used to simplify the expressions:

$$r_{jk} = -r_{kj}$$
$$t_{jk}t_{kj} = 1 - r_{jk}^2$$

(26)

In Eqs. (25), the parameters $r_{123}, t_{123}$ correspond to the total reflection and transmission coefficients, respectively. It should be noted that the denominator in all these relationships is the same.

Now, as in the previous section, we can impose the condition for the existence of guided waves, i.e., assuming that there is no incident light on the system, $A_i = 0$. Going back to Eqs. (21) and (23), we obtain:

$$A_+ = r_{21}A_- = r_{21}r_{23}A_+e^{i\phi} \Rightarrow A_+\left(1 - r_{21}r_{23}e^{i\phi}\right) = 0 \qquad (27)$$

Hence, either we have the trivial solution with all the amplitudes cancelling out, or else:

$$1 - r_{21}r_{23}e^{i\phi} = 0 \qquad (28)$$

which is the resonance condition (or eigenvalue equation) for the existence of modes in this three-media configuration. This condition is nothing more than a consistency requirement: the wave trapped in medium 2 must reproduce itself (in amplitude and phase) after each complete round trip, except for an increase in the phase in the tangential direction, x. According to Eqs. (21) and (24), the relationships between amplitudes in this circumstance are:

$$A_+ = r_{21} A_-$$
$$A_r = t_{21} A_-$$
$$A_t = t_{23} A_+ e^{i\phi/2} \quad (29)$$

Furthermore, the resonance condition of a guided mode can be equally extracted just by cancelling out the common denominators in Eq. (28). In particular, this constraint yields:

$$r_{123}, t_{123} \to \infty \quad (30)$$

Therefore, we have again singular solutions as poles of the Fresnel Coefficients. In fact, we could have chosen to write the relationships between the amplitudes of the waves in matrix form, as in the previous section. In that case, if we denote by M the matrix relating the amplitude of the incident waves with that of the transmitted and reflected waves (as in Eq. 6), the resonance condition would be obtained as det(M) = 0. I.e., the same condition we found in the former section, being the M matrix built from Eqs. (21), (23) and (24).

### 5.1 Dielectric waveguide

Let's assume that the three-media step-index planar waveguide considered in the previous section is only made of dielectric and transparent materials with real refractive indices. In this situation, the wavenumbers $k_j = n_j \omega_j / c \, (j=1,2,3)$ are real quantities and the propagation constant for radiation modes ($A_i \neq 0$) verifies $\beta < k_j$. Also, depending on the value of the refractive indices, total internal reflection at one interface, or frustrated total reflection could happen. On the other hand, if $A_i = 0$, the resonance condition can be decomposed into two different constraints:

$$|r_{21} r_{23}| = 1 \Rightarrow |r_{21}| = |r_{32}| = 1$$
$$\phi + \delta_{21} + \delta_{23} = 2m\pi \quad m \in \mathbb{Z} \quad (31)$$

where we put $r_{jk} = |r_{jk}| e^{i\delta_{jk}}$ and assume that the phase change related to the optical path, $\phi$, is real. The first relation implies there is total reflection at the two boundaries, while the second one is a constructive interference condition in medium 2. This last statement means that the phase acquired by the wave on a complete roundtrip is a multiple of $2\pi$ for whatever m-order mode. These well-known results are straightforwardly obtained here using the Fresnel formalism.

The total reflection condition implies that $n_2 > n_1, n_3$, and that the angle of propagation of waves in medium 2 is greater than the critical angle for the surrounding boundaries: $\theta_2 > \theta_{c2j}$ (j = 1,3). Moreover, under total reflection conditions, the waves in mediums 1 and 3 are evanescent waves, so the propagation constants of the modes verify: $(n_1, n_3) \omega / c < \beta_m < n_2 \omega / c$. Additionally, the specific values of $\beta_m$ and their number are obtained from the interference condition in Eq. (31).

### 5.2 Metallic waveguide

In this other case, mediums 1 and 3 are transparent dielectric materials while medium 2 corresponds to an absorbing metallic layer. As the refractive index of medium 2 is complex ($n_2 = n_2' + i n_2''$), so is the transverse component of the wave vector in this medium:

$$k_{\pm z} = \pm\sqrt{(\omega n_2/c)^2 - \beta^2}$$
$$= \pm\sqrt{(n_2'^2 - n_2''^2)(\omega/c)^2 - \beta^2 - n_2' n_2''(\omega/c)^2 i} = \pm(k'_{\pm z} + k''_{\pm z} i) \quad (32)$$

and the phase in Eq. (33) is also complex:

$$\phi = \phi' + i\phi'' = 2d(k'_{+z} + k''_{+z} i) \quad (33)$$

Hence, we decompose the resonance condition into two components, as we did in Eq. (31), obtaining:

$$|r_{21}||r_{23}|e^{-2k''_{+z}d} = 1$$
$$2k'_{+z}d + \delta_{21} + \delta_{23} = 2m\pi \quad m \in \mathbb{Z} \quad (34)$$

The major difference with respect to a dielectric guide is that the modulus of the reflectivity is below one. In consequence, in Eq. (32), the imaginary part of the transverse component of the wave vector, $k''_{+z}$, must be negative for verifying the resonance condition. Moreover, the possible values of the propagation constants, which are in general complex, are extracted from the solutions of Eq. (34) using Eq. (32). Note that in dielectric transparent media such as medium 1 or medium 3, the assumption of a complex propagation constant with a non-negligible imaginary component may be hardly correct.

A particularly simple and ideal case of metallic waveguiding is that which satisfies the condition $k'_{\pm z} = 0$. In this case, $n_2' = 0$, the permittivity $\varepsilon_2$ is real since $\varepsilon_2 = (in_2''^2)^2 = -n_2''^2 < 0$, just as in Section 4, and $k''_{+z}$ is negative. For simplicity we will address the symmetric case: $n_3 = n_1$. Following Ref. [12], we define the magnitude $R = (\alpha_2 n_1'^2 / n_2''^2 k''_{1z})$, being $\alpha_2 = -k''_{+z} > 0$ and $k_{1z}$ pure imaginary, which yields $\beta > n_1 \omega/c$. Note that $k_{1z}$ cannot be real since that would lead to $|r_{21}| = 1$, and, therefore, $k_{+z} = k''_{+z} i = 0$. Using $R$, Eq. (35) can be rewritten as:

$$\left(\frac{1-R}{1+R}\right)^2 = e^{-2\alpha_2 d}$$
$$\delta_{21} = \delta_{23} = m\pi \quad (35)$$

This expression has two different solutions, one corresponding to $\delta_{21} = 0$ ($R < 1$), and the other to $\delta_{21} = \pi$ ($R > 1$). These solutions can be summarized as:

$$\frac{1-R}{1+R} = \pm e^{-\alpha_2 d} \quad (36)$$

which yields two possible values for the propagation constant, $\beta$, one corresponding to the symmetrical or even propagation mode and the other to the anti-symmetrical or odd propagation mode, see Fig. 6. When $\alpha_2 d \gg 1$, $e^{-\alpha_2 d} \to 0$ and $R = 1$, each one of these modes is characterized by two degenerated solutions which share the same propagation constant and correspond to two SPP propagating away one from the other at the boundaries z = 0 and z = d. For the symmetric solution the SPP waves propagate in phase, while in the anti-symmetric solution they travel with opposite phases.

Finally, we note that Eq. (36) can also be written in a more usual way [13] as:

$$R^m = \tanh(\alpha_2 d/2), \quad m = -1, 1 \quad (37)$$

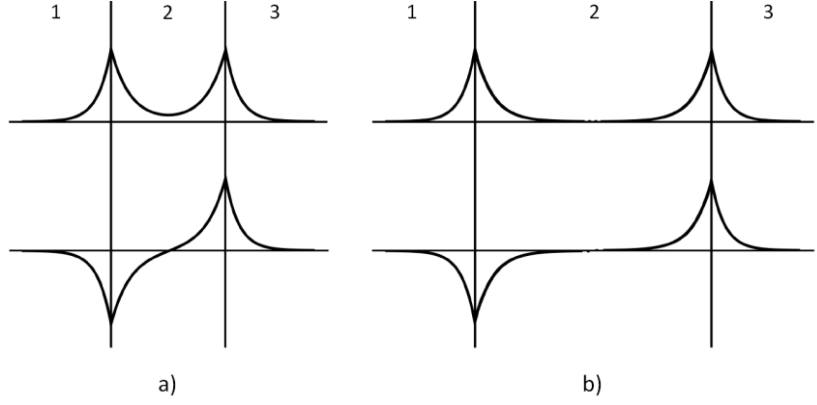

a)  b)

Figure 6. (a) symmetric and anti-symmetric guided modes of the metallic film waveguide; (b) is as (a) but with $\alpha_2 d \gg 1$.

## 6. Gradient-index waveguides

The last situation we will deal with is that of a transparent dielectric medium of variable refractive index in the z-direction which tends to a constant value at large distances. This medium satisfies the conditions (see Fig. 7):

$$n(z) = n_\pm \quad z \to \pm\infty$$
$$\max(n) > n_\pm \tag{38}$$

being the second one necessary in order to have a medium able to host guided modes.

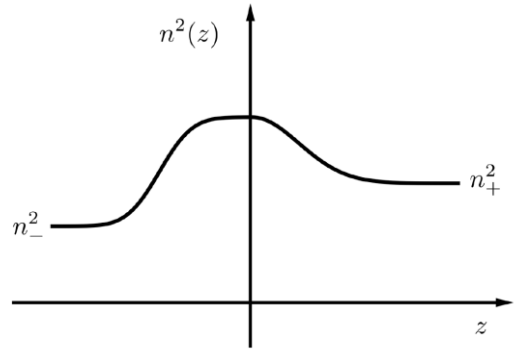

Figure 7. Typical refractive index profile for one-dimensional optical waveguiding

Within the Fresnel formalism, the Fresnel Coefficients are just defined for $z \to \pm\infty$. Considering a wave incident from the left, for any solution for the electric field, $\vec{E}(x,z)$, it must be verified at large distances that:

$$E(x,z) = \left(A_i e^{ik_- z}\vec{e}_i + A_r e^{-ik_- z}\vec{e}_r\right)e^{i\beta x} \quad z \to -\infty$$
$$E(x,z) = A_t \vec{e}_t e^{ik_+ z} e^{i\beta x} \quad z \to +\infty \tag{39}$$
$$k_\pm = \sqrt{n_\pm^2 (\omega/c)^2 - \beta^2}$$

where $\vec{e}_k \, (k=i,r,t)$ are unitary vectors that determine the vibration direction of the fields. Following the same nomenclature as in previous sections, these vectors coincide with the y-axis for TE waves, while, for TM waves, they are within the xz plane pointing to different directions except for the special case given by normal incidence. The ratio of the amplitudes of the reflected or transmitted waves with respect to that of the incident wave in Eq. (7) defines, respectively, the Fresnel Coefficients for reflection and transmission. In order to obtain those coefficients, the solutions to the Maxwell's Equations with electric fields presenting asymptotes of the functional form given by Eq. (40) must be found. Again, these solutions are different for TE and TM waves [14]:

$$\frac{d^2\varphi(z)}{dz^2} + \left[ n^2(z)\left(\frac{\omega}{c}\right)^2 - \beta^2 \right]\varphi(z) = 0 \quad E_y(x,z) = \varphi(z)e^{i\beta x} \quad (TE)$$

$$\frac{d^2\varphi(z)}{dz^2} + \frac{d\ln(n^2)}{dz}\frac{d\varphi(z)}{dz} + \left[ n^2(z)\left(\frac{\omega}{c}\right)^2 - \beta^2 \right]\varphi(z) = 0 \quad H_y(x,z) = \varphi(z)e^{i\beta x} \quad (TM)$$

(40)

being for the TM wave, the electric field solution:

$$\vec{E}(x,z) = \frac{1}{\omega n^2}\left(\frac{\partial H_y}{\partial z}\vec{x} - \beta H_y \vec{z}\right)$$

(41)

As previously, the waveguiding condition is equivalent to assume $A_i = 0$ in Eq. (39). Hence, in this case the electric field must verify:

$$E(x,z) = A_r \vec{e}_r e^{-ik_- z} e^{i\beta x} \quad z \to -\infty$$
$$E(x,z) = A_t \vec{e}_t e^{ik_+ z} e^{i\beta x} \quad z \to +\infty$$
$$k_\pm = n_\pm \omega/c$$

(42)

and the wave must verify some resonance condition. As an example, we are going to consider a TE wave in a waveguide with a symmetric exponential refractive index profile. I.e.:

$$n^2(z) = n_0^2 + \nu^2 e^{-|z|/a}$$

(43)

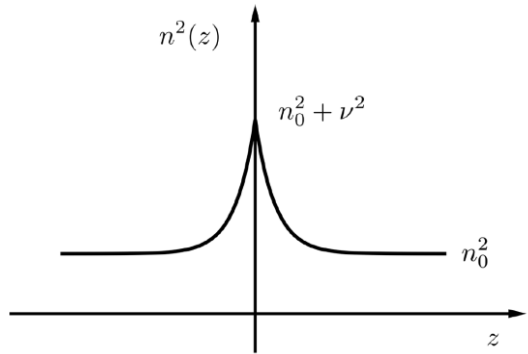

Figure 8. Symmetric exponential refractive index profile given by Eq. (43).

For this profile, the solution of Eq. (40) for the TE modes can be obtained by making the change $\zeta = 2a\nu\omega/c\, e^{-|z|/2a}$ [14,15]. Hence, it takes the form:

$$\zeta^2 \frac{d^2\varphi(\zeta)}{d\zeta^2} + \zeta\frac{d\varphi(\zeta)}{d\zeta} + \left[\zeta^2 - (\pm 2ika)^2\right]\varphi(\zeta) = 0$$

$$k^2 = n_0^2\left(\frac{\omega}{c}\right)^2 - \beta^2$$

(44)

which is nothing more than a Bessel Equation of order ±2ika. Particular solutions to this equation are the first kind Bessel functions of order ±2ika, $J_{\pm 2ika}(\zeta)$. Their behavior at $z \to \pm\infty$, ($\zeta \to 0$), is [16]:

$$J_\alpha(u) \to \frac{1}{\Gamma(\alpha+1)}\left(\frac{u}{2}\right)^\alpha \Rightarrow J_{\pm 2ika}(z) \to \frac{v^{\pm 2ika}}{\Gamma(\pm 2ika+1)}e^{\mp ik|z|} \tag{45}$$

The general solution to Eq. (44) corresponds to an arbitrary combination of first kind Bessel functions, $J_{\pm 2ika}(\zeta)$, that can be written as:

$$\varphi(\zeta) = \begin{cases} A_+ J_{2ika}(\zeta) + B_- J_{-2ika}(\zeta) & z < 0 \\ A_- J_{2ika}(\zeta) + B_+ J_{-2ika}(\zeta) & z \geq 0 \end{cases} \tag{46}$$

which according to Eq. (45) for $\zeta \to 0$ tends to:

$$\varphi(z) = \begin{cases} A_+ \dfrac{(\zeta_0/2)^{2ika}}{\Gamma(2ika+1)}e^{ikz} + B_- \dfrac{(\zeta_0/2)^{-2ika}}{\Gamma(-2ika+1)}e^{-ikz} & z \to -\infty \\ A_- \dfrac{(\zeta_0/2)^{2ika}}{\Gamma(2ika+1)}e^{-ikz} + B_+ \dfrac{(\zeta_0/2)^{-2ika}}{\Gamma(-2ika+1)}e^{ikz} & z \to +\infty \end{cases} \tag{47}$$

with $\zeta_0 = \zeta(0) = 2av\omega/c$. This solution contains four waves, two of them, those multiplied by the A constants, are "incoming" to the origin of the z coordinate and the other two, those multiplied by the B constants, are "leaving" that plane. Since we are just interested in the solutions presenting the asymptotes given by Eq. (39), the constant $A_-$ must be zero ($A_- = 0$) because only one wave is expected to move towards the center of the coordinate system, and it does that from z < 0 (see Eq. 39). Hence, taking this fact into account, Eq. (47) yields:

$$\varphi(\zeta) = \begin{cases} A_+ J_{2ika}(\zeta) + B_- J_{-2ika}(\zeta) & z < 0 \\ B_+ J_{-2ika}(\zeta) & z \geq 0 \end{cases} \tag{48}$$

Henceforth, from Eq. (39), the amplitudes of the incident, transmitted and reflected waves in the asymptotes are:

$$A_i = A_+ \frac{(\zeta_0/2)^{2ika}}{\Gamma(2ika+1)}; \quad A_r = B_- \frac{(\zeta_0/2)^{-2ika}}{\Gamma(-2ika+1)}; \quad A_t = B_+ \frac{(\zeta_0/2)^{-2ika}}{\Gamma(-2ika+1)} \tag{49}$$

Assuming that the solutions and their derivatives are continuous at z = 0 (because of the continuity of the electric field and its derivatives in z = 0 for TE waves), we can write:

$$\begin{aligned} A_+ J_{2ika}(\zeta_0) + B_- J_{-2ika}(\zeta_0) &= B_+ J_{-2ika}(\zeta_0) \\ A_+ J'_{2ika}(\zeta_0) + B_- J'_{-2ika}(\zeta_0) &= -B_+ J'_{-2ika}(\zeta_0) \end{aligned} \tag{50}$$

where the apostrophe means the derivative with respect to ζ. Eq. (41) can be written in matrix form as:

$$\begin{pmatrix} J_{2ika}(\zeta_0) \\ -J'_{2ika}(\zeta_0) \end{pmatrix} A_+ = \begin{pmatrix} -J_{-2ika}(\zeta_0) & J_{-2ika}(\zeta_0) \\ J'_{-2ika}(\zeta_0) & J'_{-2ika}(\zeta_0) \end{pmatrix} \begin{pmatrix} B_- \\ B_+ \end{pmatrix} = M \begin{pmatrix} B_- \\ B_+ \end{pmatrix} \tag{51}$$

or:

$$\zeta_0^{-4ika} \frac{\Gamma(1+2ika)}{\Gamma(1-2ika)} \begin{pmatrix} J_{2ika}(\zeta_0) \\ -J'_{2ika}(\zeta_0) \end{pmatrix} A_i = M \begin{pmatrix} A_r \\ A_t \end{pmatrix} \quad (52)$$

At this point, from Eq. (52), the Fresnel Coefficients are obtained as previously by taking the inverse of matrix M:

$$\begin{pmatrix} r \\ t \end{pmatrix} = \zeta_0^{-4ika} \frac{\Gamma(2ika+1)}{\Gamma(-2ika+1)} M^{-1} \begin{pmatrix} J_{2ika}(2v) \\ -J'_{2ika}(2v) \end{pmatrix} \quad (53)$$

If we now restrict ourselves to guided modes by taking $A_i = 0$ in Eq. (52) and we take the singular solution corresponding to the resonance condition, we obtain:

$$\det(M) = 0 \Rightarrow J_{-2ika}(\zeta_0) J'_{-2ika}(\zeta_0) = 0 \quad (54)$$

Hence, the conditions for having guided modes are $J_{-2ika}(\zeta_0) = 0$ or $J'_{-2ika}(\zeta_0) = 0$. In the first case, we have from Eq. (50) $B_+ = -B_-$, which corresponds to an anti-symmetric mode, while for the second one, we have $B_+ = B_-$, which is a symmetric mode. Moreover, it can be demonstrated that the real and imaginary parts of the first kind Bessel functions of purely imaginary order are real functions if their argument is a positive real number [17, 18]. And, even more, these functions have an infinite number of simple interspersed zeros. Thus, in order to have $\zeta$ real and positive, $k$ must be purely imaginary: $k = ik''$, $k'' \in \Re$ and, in order to have a bounded solution tending to zero at $z \to \pm\infty$, $k''$ must be also positive. Summarizing, regardless constants, we have the next solutions:

$$\varphi(z) = J_{2k''a}\left(\zeta_0 e^{-|z|/2a}\right); \quad J'_{2k''a}(\zeta_0) = 0 \quad \text{(even mode)}$$
$$\varphi(z) = \text{sgn}(z) J_{2k''a}\left(\zeta_0 e^{-|z|/2a}\right); \quad J_{2k''a}(\zeta_0) = 0 \quad \text{(odd mode)} \quad (55)$$
$$k'' = \sqrt{\beta^2 - n_0^2(\omega/c)^2} > 0$$

being the possible values of the propagation constant $\beta$ obtained from the resonance condition, Eq. (54). Again, this condition corresponds to a singular solution of a scattering problem given by Eq. (52) which implies $r, t \to \infty$.

We would like to conclude this section by indicating that, as in the previous sections, the resonance condition is twofold, or rather, it can be divided into two components. On the one hand, the resonance condition implies that the order of the Bessel function associated with a guided mode corresponds to a real number, so that the mode evanesces for large distances. On the other hand, it is shown that the Bessel function must have a zero for a particular value of its argument, which corresponds to a phase condition, as shown for example in Ref. [19].

## 7. Conclusions

In this manuscript we unify the physical treatment of one-dimensional coherent optical scattering and light-guiding problems, based on the Fresnel model of light propagation through a plane boundary between two media of different refractive indices. There is a physical distinction between these two types of problems: in a scattering problem an "external" incident wave is partially split into a reflected wave and a transmitted or refracted wave, while in a waveguide a wave can remain confined in a limited region without an external wave if a certain resonance condition is verified. Therefore, the guiding problem can be approached by first considering the dispersion problem with an external wave and then cancelling it. Considering different configurations (a single or several boundaries, or a continuous variation of the refractive index), we have shown that the singular solutions linked to the guided modes correspond to the poles of the Fresnel Coefficients of the associated scattering problem. Moreover, while the resonance condition obtained by the usual treatment corresponds to a phase condition, the one we have obtained can be divided into two components, one of which refers to the amplitude of the waves, whereas the other refers to the phase. For example, in the usual treatment of rays in a dielectric guide, it is assumed that the propagation angles of the waves in the guide correspond to the region of total reflection at the two boundaries, whereas with our formalism it is obtained from the resonance condition that guiding is only possible if this total reflection condition is given.

In our opinion, this way of treating two different types of problems following a common line allows to dig in the physical insight of the problems. Moreover, this way of proceeding also shows the shared characteristics of different guiding configurations.

As a closing remark, we would like to note that in the most general scattering problem there are two incoming waves, one from the left and another one from the right. However, when considering light guiding both waves cancel out. Therefore, for our purpose it is sufficient to treat the scattering problem with a single external wave, as we have done.

**Declaration of Competing Interest**

The authors declare that they have no known competing financial interests or personal relationships that could have appeared to influence the work reported in this paper.

**Acknowledgement**

This work was funded by the Xunta de Galicia and FEDER (GRC 508 ED431C 2020/10).